\documentclass[sigconf, authorversion,nonacm]{acmart}

\AtBeginDocument{%
  \providecommand\BibTeX{{%
    Bib\TeX}}}

\acmISBN{XX-X-XXXX-XXXX-X/XXXX/XX}

\usepackage{amsmath,amsfonts}
\usepackage{algorithmic}
\usepackage{booktabs}
\usepackage{subcaption}
\usepackage{graphicx}
\usepackage{hyperref}
\usepackage[capitalise, noabbrev]{cleveref}
\usepackage{textcomp}
\usepackage{xcolor}
\def\BibTeX{{\rm B\kern-.05em{\sc i\kern-.025em b}\kern-.08em
    T\kern-.1667em\lower.7ex\hbox{E}\kern-.125emX}}
\usepackage{amsthm}

\theoremstyle{definition}

\newcommand{\slimtilde}{\raise.17ex\hbox{$\scriptstyle\mathtt{\sim}$}}

\begin{document}
\thispagestyle{plain}
\pagestyle{plain}
\title{Reducing Compute Waste in LLMs through Kernel-Level DVFS}

\author{Jeffrey Spaan, Kuan-Hsun Chen, and Ana-Lucia Varbanescu}
\email{{j.p.spaan, k.h.chen, a.l.varbanescu}@utwente.nl}
\affiliation{%
  \institution{University of Twente}
  \city{Enschede}  
  \country{The Netherlands}
}

\renewcommand{\shortauthors}{Spaan et al.}

\begin{abstract}

The rapid growth of AI has fueled the expansion of accelerator- or GPU-based data centers. However, the rising operational energy consumption has emerged as a critical bottleneck and a major sustainability concern. Dynamic Voltage and Frequency Scaling (DVFS) is a well-known technique used to reduce energy consumption, and thus improve energy-efficiency, since it requires little effort and works with existing hardware. Reducing the energy consumption of training and inference of Large Language Models (LLMs) through DVFS or power capping is feasible: related work has shown energy savings can be significant, but at the cost of significant slowdowns. 

In this work, we focus on reducing \textit{waste} in LLM operations: i.e., reducing energy consumption without losing performance. We propose a fine-grained, kernel-level, DVFS approach that explores new frequency configurations, and prove these save more energy than previous, pass- or iteration-level solutions. 
For example, for a GPT-3 training run, a pass-level approach could reduce energy consumption by $2\%$ (without losing performance), while our kernel-level approach saves as much as $14.6\%$ (with a $0.6\%$ slowdown). 
We further investigate the effect of data and tensor parallelism, and show our discovered clock frequencies translate well for both.

We conclude that kernel-level DVFS is a suitable technique to reduce waste in LLM operations, providing significant energy savings with negligible slow-down.   

\end{abstract}




\maketitle

\section{Introduction}

In 2022, researchers at Meta AI outlined the first environmental impact study of AI~\cite{MLSYS2022_462211f6}. From their holistic approach -- spanning both development phases and the computing stack -- they emphasize the need for life-cycle measurement, co-design across data/algorithms/hardware, operational optimizations, and coordinated action to ensure AI’s environmental footprint will not continue to grow alongside its capabilities.

Yet, in the life-cycle of \textit{successful} AI models, the biggest environmental impact is attributed to operational energy consumption, for both training and inference. The energy consumption is also deeply tied to the size of AI models, i.e., training $2\times$ more typically results in $2\times$ longer training, and thus $2\times$ more energy usage. 
Alternatively, to avoid the slow-down for increasingly large models, more infrastructure is added/used.  
Therefore, as models continue to grow in size (reaching 1.8 trillion parameters for GPT-4~\cite{gtckeynote}), so does their energy consumption. Furthermore, although the energy-efficiency of AI accelerators and GPUs increases every year by $40\%$ and $29\%$~\cite{epoch2023trendsinmachinelearninghardware}, respectively, the amount of AI compute grows -- and is projected to continue to grow -- even faster at $4-5\times$ per year~\cite{epochepri2025aipower}. Thus, if the AI growth is to continue at a sustainable pace, this growing gap requires innovations that not only increase the energy-efficiency of next-generation hardware, but also of the current one.

In the world of accelerators, Dynamic Frequency and Voltage Scaling (DVFS) is one of the most powerful tools to reduce the energy consumption of a long, repetitive, and predictable workload. LLM training is such a workload. Prior studies on applying DVFS to LLMs have shown promising energy savings~\cite{batchdvfs}, while also requiring performance sacrifices. Such slow-downs, we argue, are likely to lower -- if not obliterate -- the chances of such DVFS methods to be adopted  in the current AI race-to-AGI landscape. However, we argue in this work that DVFS is not a fruitless endeavor, but it must be realistically applied. Therefore, we propose two key changes compared to previous work.

First, most work on DVFS for LLMs (or deep learning) has hitherto been applied on an iteration- or pass-level. With the advent of milli- or microsecond frequency switch latencies~\cite{predictdontreact}, we argue that this granularity is too coarse, and therefore misses critical energy-efficiency gains. In this paper, we present the first kernel-level application of DVFS for GPU-based LLMs.

Second, the optimization goal for much of the previous work was to minimize the Energy Delay Product (EDP)~\cite{11018306,9926317,9235036}: $t \cdot e$, where $t$ is the execution time (in seconds) and $e$ is the energy consumption (in Joules). Optimizing for EDP is optimizing for energy-efficiency, meaning it can compromise on time to save energy, and vice versa. Our work, on the other hand, argues for \textit{reducing compute-waste}. \textit{Waste} occurs when the resource consumption of a system (in this case, energy) is unnecessarily high, i.e., we can decrease energy consumption (via DVFS) and the execution time does not increase -- in other words, less resources are used, but nothing is lost. When a \textit{strict} waste-reduction policy is followed, an upper bound of $0\%$ time loss is enforced, whereas with a \textit{relaxed} waste-reduction policy, a pre-defined (usually minor) slow-down may be tolerated.

While optimizing for waste-reduction might not result in the most energy-efficient configuration, we propose this policy because time (and by extension, throughput) is often of higher importance, in practice, than energy. Current metrics either fail to capture this reality, or require the user to provide an unitless weight\footnote{For example, some have proposed adding an user-defined exponent $\alpha$ to the delay in the Energy Delay Product: ED$^\alpha$P $= t^\alpha \cdot e$.} of how important performance should be. In this paper, we show that optimizing for waste-reduction can enable optimizations that trade insignificant performance losses for significant energy gains.


The main goal of this work is to show the potential of fine-grained DVFS for reducing operational waste for LLMs. To this end, we provide an analysis of the expected gains, possibilities, and trends. 
Thus, the contributions of this work are as follows:
\begin{itemize}
    \item We introduce a new optimization goal, \textit{waste-reduction}, which aims to provide better incentives for energy-efficiency in a performance-dominated world (\cref{sec:waste}). 
    \item We propose the use of fine-grained kernel-level DVFS for LLMs, and empirically demonstrate the approach  outperforms previous, coarse-grained approaches (\cref{sec:coarse_analysis,sec:fine_analysis}).
    \item We consider training GPT-3~\cite{gpt3} as a case-study to show-case the application of our kernel-level DVFS, and demonstrate a reduction in the energy consumption of $14.6\%$, with only a $0.6\%$ loss in performance (\cref{sec:fine_analysis}).
    \item We further assess the effect of our frequency configurations on common forms of LLM parallelism (\cref{sec:data_parallelism,sec:tensor_parallelism}).
\end{itemize}
\section{Background}
Our work combines LLM operation and DVFS. Thus, this section provides a brief introduction into the technical details of these two broad topics.  

\begin{figure*}
\centering
\includegraphics[width=1.0\linewidth]{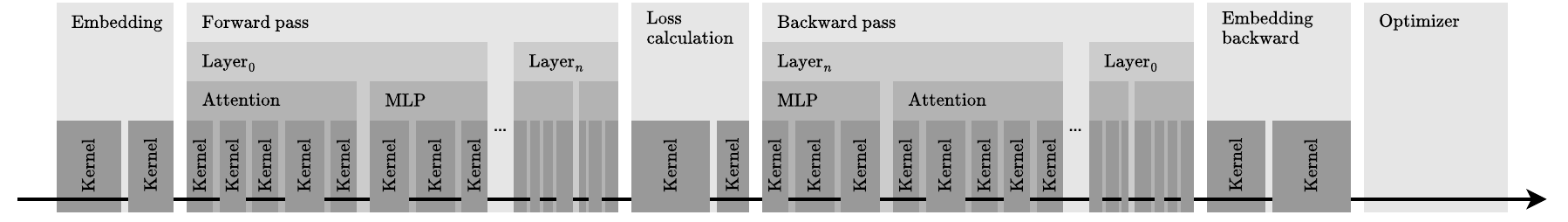}
\caption{An overview of one LLM training iteration.
}
\label{fig:llm_diagram}
\Description{}
\end{figure*}

\subsection{Large Language Models (LLMs)}

Currently, Large Language Models are the cornerstone of generative AI. Like other neural networks, they consist of a forward pass, a backward pass, and an optimizer stage. \cref{fig:llm_diagram} gives a brief overview of the components (and granularities) of LLMs. 
Each pass includes a number of transformers~\cite{vaswani2023attentionneed} \textit{layers} consisting of a self-attention block and a multi-layer perceptron (MLP). The input, consisting of a set number of tokens, flows through the network, with the output of one layer becoming the input -- the \textit{activations} -- for the next. After all layers are traversed, a new token is predicted. During training, this token is compared against the true next token and the difference, i.e., the \textit{loss}, is used in the backward pass to update the weights, using an optimizer such as Stochastic Gradient Descent (SGD) or Adam~\cite{kingma2017adammethodstochasticoptimization}. With each new training input, the model gets  gradually better at predicting the next token.

The power of transformers comes from the inclusion of a large and selective context, but the reason why transformers work as well as they do is largely unknown. However, we know that, in general, the accuracy increases with scale, both in terms of model size and amount of training data. Chasing better accuracy, we therefore need larger and faster models and, in turn, larger and faster hardware.

\paragraph{Accelerators}
Today's LLMs are trained on GPUs or AI accelerators, which are designed to handle large throughput, low communication, massively parallel workloads. Although the compute mix primarily consists of matrix multiplication operations (GEMMs), there are also other types of computations such as normalizations and data permutations. \cref{fig:fine_absolute} shows an example of the type of kernels, and their absolute execution time and energy consumption. The data clearly indicates how, depending on the input size and shape, the execution of the same kernels can vary orders of magnitude in both time and energy.

\paragraph{Parallelization}
To be able to train increasingly larger models, top-of-the-line LLMs are distributed onto multiple GPUs. For a sense of scale, LLama-3 was trained on $16000$ H100\footnote{\url{https://www.nvidia.com/en-us/data-center/h100/}} GPUs~\cite{llama3}, while llama-4 is reportedly trained on $32000$\footnote{\url{https://ai.meta.com/blog/llama-4-multimodal-intelligence/}}. To partition the model across these GPUs, ML engineers employ a variety of parallelization techniques. The most obvious type, \textit{data} parallelism (\cref{sec:data_parallelism}), divides the training data. \textit{Tensor} parallelism (\cref{sec:tensor_parallelism}) divides the tensors used in the computations -- among the hidden dimension (in the MLP) and among the heads (in the self-attention block). Lastly, \textit{pipeline} parallelism 
splits the layers of the LLM and creates multiple pipelines in which activations flow through all GPUs. Together, these approaches allow for massive parallelization, but create challenges in memory management, communication, and resource utilization.


\subsection{Dynamic Voltage and Frequency Scaling (DVFS)}
Dynamic Voltage and Frequency Scaling (DVFS) is one of the most popular methods for optimizing energy (efficiency). Although the voltage is not directly tuneable\footnote{Without voiding the warranty.}, GPU vendors such as AMD and NVIDIA allow the user to set a desired frequency within the bounds of normal operating frequencies. After applying a new frequency\footnote{In this paper, we use the terms \textit{clock}, \textit{frequency}, and \textit{clock frequency} interchangeably.} (in MHz), 
the system will run at a different speed. For example, when lowering the frequency of the processing unit, the CPU will execute fewer cycles per second. When this happens, the hardware will at the same time lower the voltage, which in turn lowers the power draw. Because the voltage scales quadratically relative to the frequency in terms of power~\cite{dvfs}, the highest clock frequencies are also the most power-hungry. 


\paragraph{Automatic DVFS}
Many different types of hardware have implemented some form of (automatic) DVFS, mainly for two scenarios: (1) to scale down clocks when idling, and (2) to scale down clocks when overheating or when drawing too much power (a.k.a. throttling). The former saves energy, while the latter prevents the system from being forced to shut down. Overall, these automatic DVFS scalers (hereafter referred to as simply the \textit{auto} clock(s)) are performance-oriented, meaning they will pursue the best performance -- through the highest clock frequencies -- so long as the GPU is not restrained by power or temperature. For NVIDIA GPUs, its exact workings are not disclosed. Furthermore, DVFS scaling behavior differs between generations and models~\cite{switchinglatency}.


\paragraph{Clock domains}
Most types of hardware feature two separate \textit{clock domains}: the \textit{core} clock and the \textit{memory} (or \textit{uncore}) clock. In many cases, lowering the \textit{memory} frequency for a \textit{compute-bound } application and lowering the \textit{core} frequency for a \textit{memory-bound} application can reduce the power consumption without affecting performance. Intuitively, this is possible because when an application is limited by a certain resource (e.g., compute capacity), one can safely reduce the use of another, unrelated, resource (e.g., memory bandwidth) until the bottleneck shifts. Finding these complementary resources and ensuring their consumption is reducing without causing a new bottleneck, is a nontrivial task. 

Furthermore, the separation between clock domains is, in practice, not so rigidly defined. For instance, the core clock domain in NVIDIA GPUs also includes the L1 and L2 caches -- making memory operations potentially affected by both clocks. These issues have led DVFS to remain a predominantly trial-and-error (auto-)tuning task~\cite{kerneltuner}.

\paragraph{Manual DVFS}
For GPUs, DVFS can be manually applied by calling a vendor-provided library or wrapper such as the NVIDIA System Management Interface\footnote{\url{https://developer.nvidia.com/system-management-interface}} (nvidia-smi) or the ROCm System Management Interface\footnote{\url{https://rocm.docs.amd.com/projects/rocm_smi_lib}} (ROCm SMI). By setting the core frequency, memory frequency, or power cap, the user can enforce a clock minimum or maximum or power draw maximum. Lowering the power cap is similar to setting clock frequencies, but no distinction is made between the memory and core clock.


Compared to CPUs, modern GPUs need much longer to process a frequency change. Sch\"one et al.~\cite{10.1145/3629526.3645040} show that an Intel Alder Lake CPU can achieve $40-100$ $\mu$s switching latencies. Huawei's Ascend NPUs can reportedly switch within $1$ ms~\cite{10.1145/3669940.3707231}. For GPUs, although unclear how fast the internal switching mechanism is, from the user's perspective, a switch can take tens of milliseconds~\cite{switchinglatency}, and switching through a tool such as \texttt{nvidia-smi} can take even longer, e.g., in our system, upwards of a $100$ ms. 

Recent work, however, has shown that with the introduction of Integrated Voltage Regulators (IVR) in (simulated) GPUs, switching latencies could be drastically improved-- (on the device side) in the order of few nanoseconds~\cite{predictdontreact}. Overall, many types of hardware have seen switching latency speedups (see \cref{sec:threats_to_validity}), making even finer-grained DVFS (on the order of hundreds or tens of instructions per frequency) within close reach.


\section{Compute Waste and its Reduction}\label{sec:waste}
To evaluate the tradeoff between energy-efficiency optimizations or optimization techniques, a metric often used, besides energy consumption itself\footnote{Which includes performance/Watt because, when comparing two optimizations using the same application, performance/Watt reduces to the inverse of the energy consumption, i.e., the difference between two optimizations becomes: $\frac{\text{FLOPs}/t_0/P_0}{\text{FLOPs}/t_1/P_1} = \frac{1/t_0/P_0}{1/t_1/P_1} = \frac{1/e_0}{1/e_1} = \frac{e_1}{e_0}$.}, is the Energy Delay Product (EDP)~\cite{edp}. EDP uses the product of the delay (in our case, execution time) $t$ (in seconds) and the energy consumption $e$ (in Joules):

\begin{equation}
    \min \quad t \cdot e
\end{equation}

A lower EDP score is considered more energy-efficient. EDP weighs time and energy equally, meaning it can sacrifice one to save another. Some have proposed squaring or cubing the time factor depending on the priority of performance~\cite{LarosIII2013}. However, as this weighting is mostly a policy decision, the original version is still widely used in research.

In this work, we propose a new, simple and more pragmatic  energy-efficiency quantifier and goal: \textit{compute waste} and \textit{compute waste reduction}, respectively\footnote{The attribute "compute" refers to the wasted resources for computing. In the paper, we often use the simplified term \textit{waste} when we refer to \textit{compute waste}.}. 

In the strict variant, we quantify waste as the difference between the current energy used and the minimum energy needed to achieve the \textit{exact} same performance. With an optimal configuration that uses $t_o$ time and $e_o$ energy, we quantify the current \textit{compute waste} as:

\begin{equation}
\begin{aligned}
    &\quad e - e_o\\     
    \text{s.t.} &\quad t_o \le t \land e_o \le e
\end{aligned}
\end{equation}

Comparing two configurations of an application (be it optimized via algorithmic choices, hardware changes, or DVFS), a lower value indicates less waste, and therefore a more energy-efficient configuration. 
In this work, we focus on quantifying the largest inefficiencies -- those who degenerately waste resources and thereby waste both time and energy. To illustrate this behavior concretely, in \cref{sec:fine_analysis}, we show an example of such a case and show how EDP discounts it.
\begin{figure}[tbh!]
\centering
\includegraphics[width=1.0\linewidth]{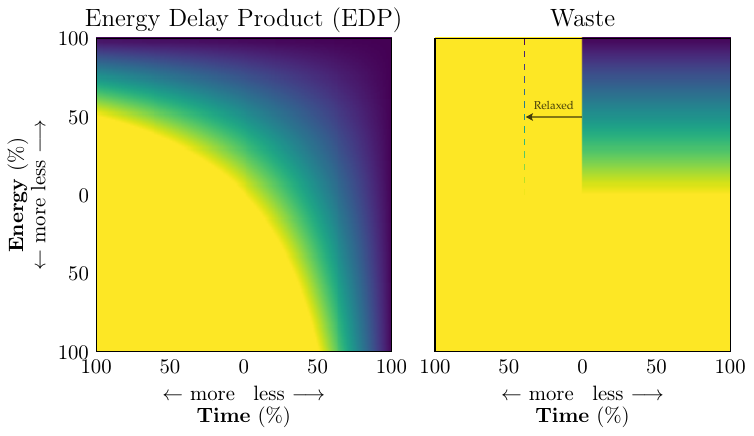}
\caption{Desirability of configurations according to the EDP and \textit{waste} optimization goals. Darker colors indicate better scores.}
\label{fig:optimization_goal_area}
\Description{}
\end{figure}
\cref{fig:optimization_goal_area} depicts the desirability scores of time and energy loss/gain pairs ranging from $-100\%$ to $+100\%$ when pursuing the Energy Delay Product (EDP) and \textit{waste} optimization goals. Points with the same color indicate equal scores. As can be seen in the left figure, EDP does not differentiate between time and energy ($2t \cdot e = t \cdot 2e$) -- nor between gain and loss ($2t \cdot \frac{1}{2}e = t \cdot e$). This culminates into, in our opinion, counter-intuitive tradeoffs. For example, according to EDP, saving $50\%$ energy without losing performance is considered as good as saving $75\%$ energy with a $100\%$ time loss.

Our approach, in the figure on the right, on the other hand, discards the left quadrant (time loss) and the bottom right quadrant (energy gain). As can also be seen, \textit{waste} does not differentiate between the scale of time savings. We do not focus on this because ``efficiency'' optimizations travel towards the top right, while ``energy-efficiency'' optimizations travel towards the top. Both trends are captured by \textit{waste}. Optimizations where the time is reduced while the energy remains equal (i.e., traveling to the right) likely increase the power consumption (since $e = t \cdot P$) and are therefore more related to general performance engineering techniques -- for which other metrics are more appropriate.

\section{Experimental Setup for Case Study}

\paragraph{Objective.} Our goal is to empirically demonstrate that applying DVFS to LLMs operation at different granularities can reduce compute waste. To demonstrate why granularity matters, we systematically apply DVFS to an LLM on two granularities. First, on a coarse-grained level (e.g., a forward pass) and then on a finer-grained kernel-level (e.g., a GEMM). Using the clock combinations discovered from the latter, we also test the effects of data parallelism and tensor parallelism. In the following paragraph, we explain how we setup the experiments needed for this empirical analysis. 

\paragraph{Model.}
We perform our experiments on the GPT-3-xl~\cite{gpt3} $1.3$ billion parameter model. The model consists of 24 layers, a hidden size of $2048$, and $16$ attention heads. Unlike GPT-3, we fix the sequence length to $1024$ to allow for experimentation with higher batch sizes (we use a default of $40$).

\paragraph{Software.}
To test the kernels of this model in isolation, we use \texttt{llm.c}\footnote{\url{https://github.com/karpathy/llm.c}}, a pure C/CUDA LLM implementation primarily designed for recreating the pre-training of GPT-2/3. \texttt{llm.c} allows us to isolate and alter individual kernels, while remaining comparable -- and competitive -- with more commonly used frameworks such as PyTorch\footnote{\url{https://pytorch.org/}} or TensorFlow\footnote{\url{https://www.tensorflow.org/}}. We have extended their implementation to support tensor and sequence parallelism (excluding communication) following the methods proposed by Megatron-LM~\cite{megatron}.

\paragraph{Hardware}
All our experiments are performed on an Nvidia RTX 3080 Ti GPU. Although this GPU only has 12GB device memory, our experiments only require one layer of the LLM -- since they are identical performance-wise, which allows us to scale the memory requirements down drastically. Furthermore, this GPU allows us to set $6$ different memory frequencies and $127$ different core frequencies between 210 and 2100 MHz with $15$ MHz increments. 
To save time, our experiments only include core clocks with 210 MHz increments. 

\paragraph{Search}
Finding the `best' frequency for a hardware-software pair can be an arduous process. Much effort has been put into performing such an automation more efficiently. State-of-the-art tools like BatchDVFS~\cite{batchdvfs} or the work by Wang et al.~\cite{10.1145/3669940.3707231}, and, more generally, KernelTuner~\cite{kerneltuner} and the work by Ali et al.~\cite{ALI202371}, use performance and power modeling or optimization techniques such as gradient descent or genetic algorithms to increase the speed of the search. Unfortunately, they do not fit our problem space, which involves multiple simultaneously optimized kernels. Furthermore, as one of our goals is to compare different optimization and aggregation strategies, we require a full exhaustive search to evaluate the tradeoffs. 
Overall, the fine-grained experiment, using all kernels and clock combinations, took around $3$ GPU days to complete, which is not insignificant, but negligible compared to the training times of popular LLMs.

\paragraph{Workflow}
To measure the time and energy of each kernel/pass - core clock - memory clock combination, we first warm up the GPU using the maximum frequencies for $100$ seconds (aiming to reach peak temperature\footnote{As we assume the training will always try to fully utilize the GPU.} of around $77$ degrees Celsius). Then, we set the desired frequencies (the actual frequencies are allowed to range between the minimum and the target frequencies) and warm up the target kernel or pass for $25$ seconds. Afterwards, we measure the kernel/pass for $5$ seconds, and repeat the last two steps for the next kernel or pass. We use CUDA events to measure time and \texttt{nvmlDeviceGetTotalEnergyConsumption} to measure energy. Our measurements have to be relatively long (compared to the true execution times) because the internal power measurement resolution can be quite low -- for our GPU, around $1$ second\footnote{When measuring the instant power (\texttt{NVML\_FI\_DEV\_POWER\_INSTANT}).}. Prior work has shown that although the internal sensor can have a significant lag, a longer measurement window can ensure measurements approach the true power consumption~\cite{powersensor, 10.1109/SC41406.2024.00028}.


\section{Pass-Level Compute-Waste Analysis }\label{sec:coarse_analysis}
\begin{figure}
\centering
\includegraphics[width=1.0\linewidth]{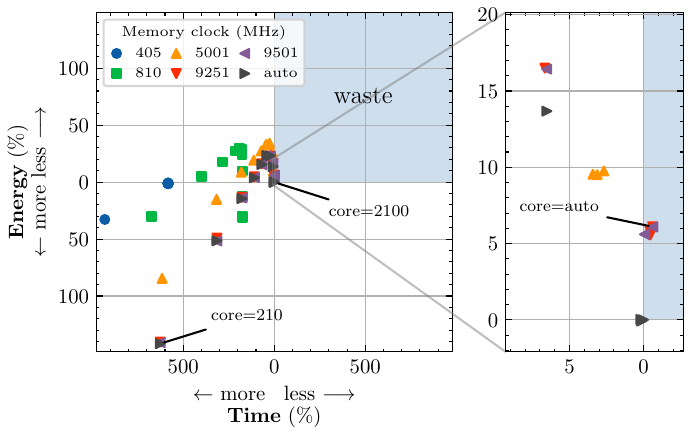}
\caption{\textbf{Forward pass}. Depicts the percentage of time and energy gained or lost when the reducing the memory clock and core clock from the (\texttt{auto}, \texttt{auto}) baseline (at the origin). }
\label{fig:coarse_forward}
\Description{}
\end{figure}

\begin{figure}
\centering
\includegraphics[width=1.0\linewidth]{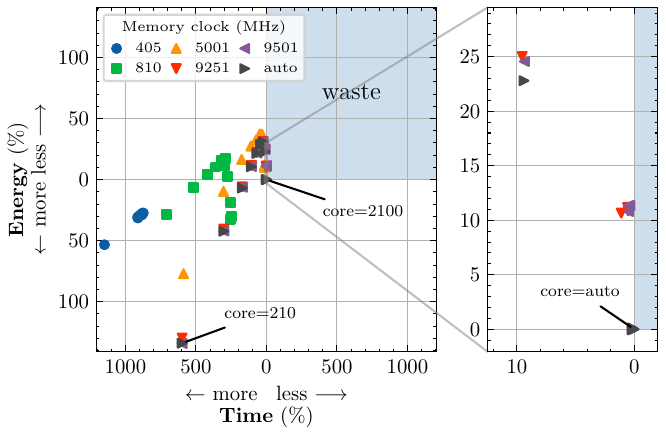}
\caption{\textbf{Backward pass}. Depicts the percentage of time and energy gained or lost when the reducing the memory clock and core clock from the (\texttt{auto}, \texttt{auto}) baseline (at the origin). }
\label{fig:coarse_backward}
\Description{}
\end{figure}

First, we investigate the potential of DVFS to reduce compute waste when applied on a coarse-grained level, meaning we apply a single core clock and memory clock frequency per \textit{pass}. The results for the forward pass and the backward pass are shown in \cref{fig:coarse_forward,fig:coarse_backward}, respectively. The figures show a blue square containing all clock configurations that are considered better than the baseline policy (i.e., using only \texttt{auto} clocks) under the \textit{waste-reduction} optimization policy. The configuration with the most energy saved (vertically, the highest) are selected as the `best' clock.

For each memory clock, the curve first shows a steep rise (small time losses, large energy savings), and then gradually declines leftwards towards increased energy consumption and time. The configurations at the top of the curve are the most energy-efficient. We observe that, the lower the memory clock, the lower-left its curve is -- meaning that, at a pass-level, only the higher memory clocks achieve the highest performance and lowest energy consumption. This happens because the compute mix in a pass-level granularity is inherently varied, with (internally) different kernels preferring (that is, running best at) different frequencies -- and different \textit{types} of clock frequencies. The \texttt{auto} clocks therefore veer to the highest frequencies, a choice that makes most (if not all) other tested frequencies lower in comparison. Since the `average' kernel prefers the highest frequencies,  all other frequencies hence (overall) perform worse than the \texttt{auto} clocks.

We further observe that the 405 MHz memory clock will only be set for core clocks less than or equal to 420 MHz. Higher core clocks will in practice raise the memory clock to 810 MHz, which is why most of its points are overlapped with 405 MHz's points in \cref{fig:coarse_forward} and \cref{fig:coarse_backward}. Regardless, both memory clocks never yield the best score in terms of time, energy, \textit{waste}, or EDP.

Six clock combinations (memory clock, core clock) fall in the \textit{waste} square for the forward pass: ($9501,\texttt{auto}$), ($9501,2100$), ($9501,1890$), ($9251,\texttt{auto}$), ($9251$,$2100$), and ($9251$,$1890$), with similar time and energy savings, roughly $0.5\%$ time and $6\%$ energy.

Although the results for both passes are similar, reducing the clock frequencies during the backward pass will result in performance loss across the board, meaning that if the strict \textit{waste-reduction} policy is followed, the original \texttt{auto} clock remains superior. However, as can be seen in \cref{fig:coarse_backward}, some configurations can reduce energy by roughly $12\%$ while only showing $<1\%$ delay. This asymptotic behavior emerges because, when lowering the frequency, the time will, in general, decrease linearly, while the power draw will decrease quadratically with the voltage (which scales linearly with the frequency\footnote{In practice, some frequencies share the same voltage, so the frequency-to-voltage curve is piecewise linear.}). As the highest frequencies are the most inefficient, a \textit{relaxed} \textit{waste-reduction} goal might be a wiser choice in this scenario.

For both passes, with a strict \textit{waste-reduction} goal, the coarse-grained approach can save $1.98\%$ of the total energy consumption and $0.2\%$ of the total execution time.

\begin{figure}
\centering
\includegraphics[width=1.0\linewidth]{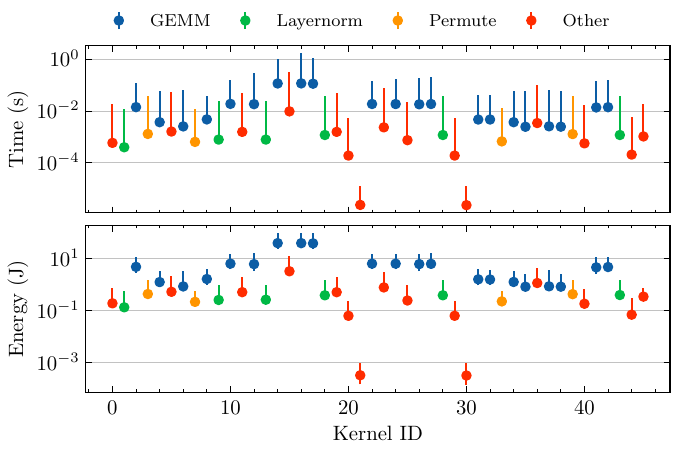}
\caption{Absolute time and energy for all kernels. Dots indicate the \texttt{auto} configuration, while the error bars indicate the minimum and maximum value achieved by \textit{any} clock configuration.}
\label{fig:fine_absolute}
\Description{}
\end{figure}

\section{Kernel-Level Compute-Waste Analysis}\label{sec:fine_analysis}
In this section, we perform the same analysis as in \cref{sec:coarse_analysis}, but we deploy DVFS at \textit{kernel-level}. \cref{fig:fine_absolute} shows an overview of all types of executed kernels. Although some of these kernels are identical, the input size and/or input dimensions might not be the same, which, besides from the absolute time and energy differences, can result in different DVFS impact. We therefore measure each kernel invocation separately. Furthermore, for kernels that are part of the transformer layer ($\#2$-$\#13$, $\#19$-$\#43$), we multiply the result by the number of layers ($24$) to simulate the time and energy of a full iteration.

\cref{tab:fine} shows the best clocks with their corresponding time and energy gain/loss result for all $46$ kernels when aiming for strict \textit{waste-reduction}. The results show that there are large differences in how much time and energy the best clocks save between kernels. For instance, kernel $\#2$, a GEMM, saves $15.41\%$ of the energy and improves performance by $2.36\%$, while kernel $\#4$, also a GEMM, saves only $2.74\%$ of the energy with a similar performance gain. Furthermore, we observe that kernels either prefer a lower memory clock or a lower core clock -- but not both.

\paragraph{Local vs. global optimization}
One might observe that \cref{tab:fine} contains clock combinations that result in time losses, something which is not allowed under a strict \textit{waste-reduction} policy. This is possible because we aggregate and optimize \textit{globally}, not \textit{locally}. The latter forces every kernel to adhere to the $0\%$ time loss rule, while the former only prohibits the total execution time from decreasing -- thus allowing virtually unlimited per-kernel time loss. This strategy allows kernels to cooperatively compensate for losses in time -- finding one global optimum\footnote{We use a constraint solver to find the optimal global allotment in near instant time.} instead of multiple local optima. As an example, assume kernel A can either save $2\%$ time and $20\%$ energy or save $3\%$ time and $15\%$ energy, while kernel B could save $35\%$ energy but with a $1\%$ time \textit{increase} or save $0\%$ for both. When aggregating locally, kernel A's first and kernel B's second configuration will be selected. When optimizing globally, one could select kernel A's second configuration to compensate the time losing first configuration of kernel B. If the absolute time and energy is the same, the local strategy will only save $1\%$ time and $10\%$ energy, while the global strategy will save $1\%$ time and $25\%$ energy.

\begin{figure}
\centering
\includegraphics[width=1.0\linewidth]{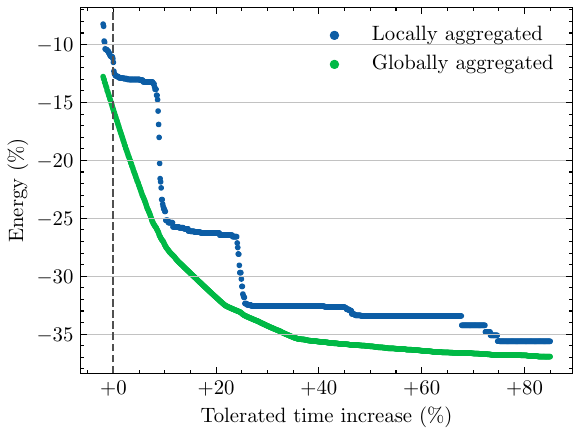}
\caption{Energy savings, optimized locally and globally, for different tolerated time increase (performance loss) thresholds. The dotted line denotes $0\%$ (strict waste). }
\label{fig:fine_threshold}
\Description{}
\end{figure}

\paragraph{Relaxed waste}
\cref{fig:fine_threshold} shows the total amount of energy saved when increasing the tolerated performance loss, with $0\%$ (strict \textit{waste-reduction}) denoted by the dotted line. Note that this is a threshold, not the actual time gain (which may be lower -- but not higher). The time gain for the global strategy almost always approaches the threshold. We can observe that a strict \textit{waste-reduction} policy achieves a $\slimtilde15\%$ energy reduction with no performance loss. The local strategy can only reduce $\slimtilde 12\%$. Indeed, the global optimization strategy, by design, always outperforms the local strategy, since the global strategy can always optimize locally, and not the other way around. For increasingly higher tolerated performance losses, the amount of extra energy reduced diminishes for both strategies. With a $30\%$ time increase, we can only save $35\%$ of the energy consumption. The figure also shows how exclusively optimizing for energy is highly wasteful; the highest energy reduction ($36.9\%$) can only be achieved with an $84\%$ performance loss.

Note that, although the points may also indicate a trend that continues (leftwards) into saving more time, this is not the case. The maximum time savings achieved by any combination of clock configurations was found to be $2\%$.

The plateaus for the local optimization strategy are likely an artifact of our range of tested core clocks, which are separated by $210$ MHz, which is $10\%$ of the maximum clock ($2100$ MHz). For many (compute-bound) kernels, every $10\%$ drop in performance allows the kernels to select the `next' $10\%$ lower core clock: $2100$, $1890$, $1680$, etc. The global strategy is able to more effectively compensate for these jumps by distributing the allowed performance loss to the kernels that reduce relatively more energy.

Although optimizing globally seems superior, the benefit of optimizing locally lies in isolation. The global strategy is only feasible because we have access to the results for all clock combinations for all kernels. In practice, an \textit{efficient} frequency search algorithm that can optimize \textit{globally} will be more complex and will require a larger search space.

\paragraph{Waste vs. EDP}
\cref{tab:compare} contains the total time and energy loss/gain for the coarse- and fine-grained approaches using the local and global optimization strategies. As shown, the fine-grained approach can save considerably more energy ($15.65\%$) than the coarse-grained approach ($2.07\%$). Furthermore, the global strategy is able to effectively find the configurations that maximize energy reduction within the performance loss threshold.

The table also shows the results when optimizing for EDP. For EDP, both granularities yield a $\slimtilde 10\%$ time loss, although the fine-grained approach is able to save $2\%$ more energy, for a total of $27.52\%$. Unlike waste, the difference between local and global optimization is negligible. Overall, while optimizing for EDP can yield larger energy savings, it also introduces significant slowdowns. Without a deeper inspection, one might conclude that energy savings naturally require a tradeoff in performance, which, as shown in the \textit{waste} column, is not always true.

\begin{figure}
\centering
\includegraphics[width=1.0\linewidth]{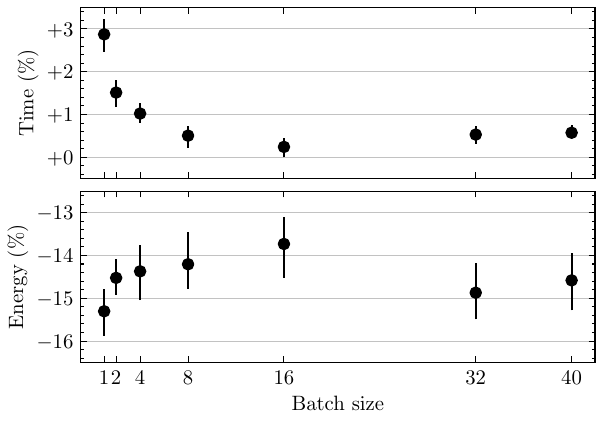}
\caption{\textbf{Data parallelism}. Time and energy gained or lost compared to the \texttt{auto} configuration using the best clocks for batch size $40$. We measured 10 repetitions using both the best and auto clocks. Error bars show the best and worst gain for all $10 \cdot 10$ scenarios.}
\label{fig:data_parallelism}
\Description{}
\end{figure}

\paragraph{Validation}
The results show our clock configurations can thus save energy without losing performance. In reality, however, we expect the performance loss to not be exactly zero due to the accumulation of measurement errors. To confirm this, we repeated the measurements using the best clocks and the \texttt{auto} clocks (as shown in \cref{tab:fine}) $10$ times, for a total of a $100$ best-to-\texttt{auto} comparisons. The results, as seen in \cref{fig:data_parallelism} at a batch size of 40, show that the mean performance loss is indeed higher (around $0.6\%$) and the mean energy savings indeed lower (around $14.6\%$). Furthermore, the error bars show that no best-clocks-\texttt{auto}-clocks combination (out of a $100$) achieved the originally discovered savings. We believe this happens because our approach naturally gravitates towards selecting (and accumulating) measurement outliers. Consequently, we expect the relative error to increase with the number of kernels. The only way to minimize this error is to increase the number (or duration) of measurements or to reduce the magnitude of measurement errors, e.g., via more accurate power measurement tools~\cite{powersensor}.


\begin{table}
\caption{Optimal clocks, execution time loss/gain, and energy consumption loss/gain per kernel, optimizing for globally aggregated strict \textit{waste}. Backward kernels are denoted by $\leftarrow$.}
\label{tab:fine}
\begin{center}
\begin{normalsize}
\begin{tabular}{rlrrrr}
\toprule
 & & Mem. & Core & Time & Energy \\
 & & (MHz) & (MHz) & (\%) & (\%) \\
\midrule
\multicolumn{5}{@{}l}{\textit{Embedding}} \\
$0$ & WTE \& WPE & $\texttt{auto}$ & $630$ & $+0.32$ & $-33.01$ \\
$1$ & Layernorm & $\texttt{auto}$ & $1050$ & $+0.77$ & $-29.20$ \\
\midrule
\multicolumn{5}{@{}l}{\textit{Forward} $... \times \text{\#layers}$} \\
$2$ & GEMM & $5001$ & $\texttt{auto}$ & $-2.36$ & $-15.41$ \\
$3$ & Permute & $9501$ & $1680$ & $+1.52$ & $-10.83$ \\
$4$ & GEMM & $9501$ & $\texttt{auto}$ & $-1.78$ & $-2.74$ \\
$5$ & Softmax & $9501$ & $1050$ & $-0.03$ & $-11.97$ \\
$6$ & GEMM & $9251$ & $\texttt{auto}$ & $-1.27$ & $-4.55$ \\
$7$ & Permute & $9251$ & $\texttt{auto}$ & $-1.42$ & $-5.68$ \\
$8$ & GEMM & $5001$ & $\texttt{auto}$ & $-2.08$ & $-14.54$ \\
$9$ & Residual & $\texttt{auto}$ & $840$ & $+0.59$ & $-30.97$ \\
$10$ & GEMM & $5001$ & $\texttt{auto}$ & $-2.67$ & $-15.21$ \\
$11$ & GELU & $9501$ & $630$ & $+0.03$ & $-33.21$ \\
$12$ & GEMM & $5001$ & $\texttt{auto}$ & $-3.02$ & $-13.77$ \\
$13$ & Residual & $9501$ & $1050$ & $+0.43$ & $-32.49$ \\
\midrule
\multicolumn{5}{@{}l}{\textit{Loss calculation}} \\
$14$ & GEMM & $5001$ & $\texttt{auto}$ & $-2.60$ & $-15.72$ \\
$15$ & Softmax & $9501$ & $1680$ & $+1.98$ & $-26.65$ \\
$16$ & GEMM & $9251$ & $\texttt{auto}$ & $-0.96$ & $-7.75$ \\
$17$ & GEMM & $5001$ & $1680$ & $+8.98$ & $-29.31$ \\
$18$ & $\leftarrow$ Layernorm & $\texttt{auto}$ & $1260$ & $+1.92$ & $-29.05$ \\
\midrule
\multicolumn{5}{@{}l}{\textit{Backward} $... \times \text{\#layers}$} \\
$19$ & GELU & $9501$ & $630$ & $+0.03$ & $-33.14$ \\
$20$ & $\leftarrow$ Bias & $\texttt{auto}$ & $1260$ & $+0.88$ & $-31.87$ \\
$21$ & $\leftarrow$ Bias reduce & $\texttt{auto}$ & $\texttt{auto}$ & $+0.00$ & $+0.00$ \\
$22$ & GEMM & $5001$ & $\texttt{auto}$ & $-2.73$ & $-15.36$ \\
$23$ & $\leftarrow$ GELU & $9501$ & $840$ & $-0.04$ & $-26.88$ \\
$24$ & GEMM & $5001$ & $1680$ & $+10.13$ & $-30.80$ \\
$25$ & $\leftarrow$ Bias & $\texttt{auto}$ & $1050$ & $+0.42$ & $-31.34$ \\
$26$ & GEMM & $5001$ & $\texttt{auto}$ & $-2.68$ & $-13.30$ \\
$27$ & GEMM & $9251$ & $\texttt{auto}$ & $-1.65$ & $-6.77$ \\
$28$ & $\leftarrow$ Layernorm & $\texttt{auto}$ & $1260$ & $+1.89$ & $-29.42$ \\
$29$ & $\leftarrow$ Bias & $9501$ & $1260$ & $+0.88$ & $-32.68$ \\
$30$ & $\leftarrow$ Bias reduce & $\texttt{auto}$ & $\texttt{auto}$ & $+0.00$ & $+0.00$ \\
$31$ & GEMM & $5001$ & $\texttt{auto}$ & $-2.46$ & $-14.19$ \\
$32$ & GEMM & $5001$ & $\texttt{auto}$ & $-2.08$ & $-12.42$ \\
$33$ & Permute & $9501$ & $\texttt{auto}$ & $-0.31$ & $-5.99$ \\
$34$ & GEMM & $9501$ & $\texttt{auto}$ & $-1.85$ & $-2.70$ \\
$35$ & GEMM & $9251$ & $\texttt{auto}$ & $-0.67$ & $-6.11$ \\
$36$ & $\leftarrow$ Softmax & $9501$ & $\texttt{auto}$ & $-0.17$ & $-5.23$ \\
$37$ & GEMM & $9251$ & $\texttt{auto}$ & $-1.52$ & $-3.51$ \\
$38$ & GEMM & $9501$ & $\texttt{auto}$ & $-0.53$ & $-5.55$ \\
$39$ & Permute & $9501$ & $1470$ & $+2.62$ & $-18.35$ \\
$40$ & $\leftarrow$ Bias & $\texttt{auto}$ & $1260$ & $+0.60$ & $-30.72$ \\
$41$ & GEMM & $5001$ & $1680$ & $+9.03$ & $-29.34$ \\
$42$ & GEMM & $9501$ & $\texttt{auto}$ & $-1.72$ & $-6.77$ \\
$43$ & $\leftarrow$ Layernorm & $9501$ & $1260$ & $+1.86$ & $-30.49$ \\
\midrule
\multicolumn{5}{@{}l}{\textit{Embedding backward}} \\
$44$ & $\leftarrow$ WPE & $9501$ & $1260$ & $+2.37$ & $-31.35$ \\
$45$ & $\leftarrow$ WTE & $\texttt{auto}$ & $1680$ & $+7.25$ & $-28.37$ \\
\bottomrule
\end{tabular}
\end{normalsize}
\end{center}
\end{table}

\begin{table}[htp]
\caption{Total time and energy gains/losses for different optimization goals and aggregation levels and techniques.}
\label{tab:compare}
\begin{center}
\begin{small}
\begin{tabular}{lrrlrr}
\toprule
Optimize for & \multicolumn{2}{c}{EDP} & & \multicolumn{2}{c}{\textit{Waste}} \\
\cmidrule{2-3}
\cmidrule{5-6}
 & Time & Energy & & Time & Energy \\
 & (\%) & (\%) & & (\%) & (\%) \\
\midrule
\multicolumn{1}{@{}l}{Coarse-grained} & & & \\
Local optima & $+10.21$ & $-25.42$ & & $-0.20$ & $-1.98$ \\
Global optimum & $+10.21$ & $-25.42$ & & $-0.10$ & $-2.07$ \\
\midrule
\multicolumn{1}{@{}l}{Fine-grained} & & & \\
Local optima & $+10.03$ & $-27.34$ & & $-1.78$ & $-11.54$ \\
Global optimum & $+10.28$ & $-27.52$ & & $+0.00$ & $-15.64$ \\
\bottomrule
\end{tabular}
\end{small}
\end{center}
\end{table}



\section{Data parallelism}\label{sec:data_parallelism}
All multi-GPU LLMs use data parallelism to distribute the training work across GPUs. Data parallelism works by duplicating the LLM -- including all weights and optimizer states -- across GPUs and by dividing the batch size -- the number of simultaneously evaluated input samples. Although this technique requires a very large \textit{total} batch size, research has shown that the models still converge, albeit at a slower pace~\cite{keskar2017largebatchtrainingdeeplearning}.

To test the effect of the per-GPU batch size on our discovered clock configurations (from \cref{tab:fine}), we measure the time and energy loss/gain compared to the \texttt{auto} clock for all kernels while roughly halving our baseline batch size (40) until only one batch is executed.

\cref{fig:data_parallelism} shows the gains and losses for all batch sizes. Although the absolute percentage point differences are small, lower batch sizes gain relatively more time ($3\%$ vs. $0.6\%$) and save relatively more energy ($15.3\%$ vs $14.6\%$). This points to smaller batch sizes preferring higher clocks -- since lower clocks generally save more energy but gain more time. Because lowering the batch sizes lowers utilization, a possible result is that the GPU throttles less, i.e., it can sustain higher clocks because the power draw or the temperature decreases.

\cref{fig:data_parallelism} also shows how the execution time gains per batch size are far more consistent than the energy savings. Considering that the energy is simply the time multiplied by the power, it shows that the variability in our measurements is mostly caused by the latter.



\section{Tensor parallelism}\label{sec:tensor_parallelism}
Tensor (or model) parallelism~\cite{megatron} is a parallelization technique used to (further) divide the workload among GPUs by slicing the tensors of the matrix computations (in the hidden dimension), performing each submatrix operation on a different GPU, and aggregating the result. Like data parallelism, the use of tensor parallelism is often born out of necessity -- to accommodate a desired model size -- not in the pursuit of performance. Therefore, where in our experiments a higher tensor parallelism degree results in lower GPU utilization, in practice, this is often compensated by a larger model size -- through one of many dimensions.

Besides tensor parallelism we have also implemented \textit{sequence} parallelism which can further divide the previously unparallelized elements such as the encoder, the layer normalizations, and the loss calculation. To accomplish this, the tensors are divided in the sequence dimension.

The maximum tensor parallelism degree is limited to the number of heads (in the self-attention mechanism) which in our model is $16$. Furthermore, because of high -- and frequent -- communication requirements, tensor parallelism is often restricted to one node (typically containing $8$ or at most $16$ GPUs). We therefore only test tensor parallelism degrees up to $16$.

\begin{figure}[t]
\centering
\includegraphics[width=1.0\linewidth]{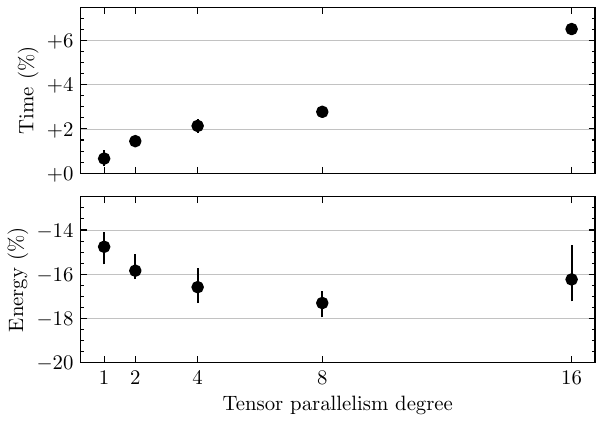}
\caption{\textbf{Tensor parallelism}. Time and energy gained or lost compared to the \texttt{auto} configuration using the best clocks for tensor parallelism degree $1$. We measured $10$ repetitions using both the best and auto clocks. Error bars show the best and worst gain for all $10 \cdot 10$ scenarios.}
\label{fig:tensor_parallelism}
\Description{}
\end{figure}

\cref{fig:tensor_parallelism} shows the time and energy loss/gain compared to the \texttt{auto} clock for all kernels using our discovered clock configurations (from \cref{tab:fine}) while doubling the tensor parallelism degree (starting at $1$, i.e., no tensor parallelism). Similar to the batch size, the results show the gains and losses diverge when increasing the tensor parallelism degree. An outlier is a degree of $16$, which gains more than twice the time gain of a degree of $8$ ($6.5\%$ vs. $2.7\%$) while saving relatively less energy ($16.2\%$ vs. $17.3\%$). 

We also observe that, compared to the data parallelism experiment, the time gains and energy savings are higher for a similar reduction in workload size. For example, for a degree of $4$, the workload is roughly $4$ times smaller which should make it similar to a batch size of $\frac{40}{4} \approx 8$. However, the time gain and energy savings are both two percentage points higher: $2.1\%$ vs. $0.5\%$, and $16.6\%$ vs. $14.2\%$, respectively. Furthermore, the absolute time and energy consumption also differ starkly: $1.12$ s and $325$ J for a batch size of $8$ and $1.8$ s and $503$ J for a tensor parallelism degree of $4$, almost twice as much. Thus, it seems the total workload size alone is poor indicator of performance, energy, or how well the discovered clock frequencies will translate to different parallelization methods.



\section{Discussion}\label{sec:threats_to_validity}
In previous sections, we have shown that DVFS can lower the GPU energy consumption of LLM training significantly with, when optimizing for \textit{waste-reduction}, almost no performance loss. On the implementation side however, DVFS brings it own set of challenges. In this section, we discuss DVFS's challenges, as well as its benefits.

\paragraph{Error accumulation}
As stated in \cref{sec:fine_analysis}, our fine-grained approach has a tendency to select and accumulate the clocks that produce (positive) measurement outliers. When repeating the experiments however, we found the total losses/gains differences to be contained to $\slimtilde 0.5$ percentage point.

\paragraph{GPU heterogeneity}
The effectiveness of DVFS depends on the GPU and the generation of GPU~\cite{11018306}. To confirm this for our results, we have performed the same fine-grained experiment (\cref{sec:fine_analysis}) with the same model and implementation on an A4000. We could not directly re-apply our discovered frequencies to this GPU because the range of applicable core and memory clocks is different. When rerunning the experiment, the A4000 achieved a $9.56\%$ energy reduction with $0\%$ performance loss. Compared to our previous analysis, the results show kernels prefer the same clock types, but reduce the clocks less aggressively. When optimizing for EDP, the results are again lower, with an $8.28\%$ reduction in energy and $2.33\%$ reduction in time. Thus, to improve energy-efficiency, besides looking at a GPU's initial/peak energy-efficiency, one should also look at its DVFS potential. To overcome heterogeneity between different GPUs of the same type, we expect differences can be counteracted by remeasuring a subset of the clocks surrounding the optimal clocks on each individual GPU.

\paragraph{Frequency switching latency}
Because GPU frequency switching latencies might overshadow the latency of particular kernels, not all of our discovered frequencies can currently be (successively) applied without overlap -- which our results show worsens the DVFS potential. Although the latency is currently (too) high, Velicka et al.~\cite{switchinglatency} demonstrate newer generation GPUs achieve lower latencies: from $\slimtilde 150$ ms using an RTX Quadro 6000 (Turing) to $\slimtilde 6$ ms using a H200 (Hopper). Whether this trend continues is unknown. Other types of hardware are currently able to decrease the latency further. Huawei's Neural Processing Units (NPUs) can achieve $1$ ms frequency switching latencies~\cite{10.1145/3669940.3707231}, while, according to AMD, its 2018 Raven Ridge APU can achieve $1$ $\mu$s~\cite{amdraven}.

\paragraph{Reliability}
DVFS can improve reliability because lower clocks consume less power, less power leads to lower temperatures, and lower temperatures leads to fewer failures~\cite{1541878}. Furthermore, DVFS can potentially benefit aging because lower clocks require lower voltages, and lower voltages slow down transistor degradation.

\section{Related work}

\paragraph{DVFS fundamentals}
In 2005, Cameron et al. outlined the first review of power-aware distributed computing~\cite{1541878}. In this work, they promote the idea of optimizing for energy-efficiency without affecting performance -- but note that the level of performance sacrifice (and thereby the desired energy savings) is ultimately up to the user. Through their distributed power profiling and management framework (PowerPack), they show DVFS and energy-aware scheduling can be used to achieve near performance-loss-less energy savings. They futher note that ``performance is a hard constraint and power is a soft constraint in high-performance computing''. We argue the same is true for AI today, insofar that power limitations are simply mitigated by scaling the amount of compute.

Eyerman and Eeckhout present the first fine-grained automatic CPU DVFS scaler prototype using mechanistic performance modeling~\cite{10.1145/1952998.1952999}. Their approach scales down the voltage and frequency upon individual off-chip memory accesses, with a nanosecond granularity (20 processor cycles). Although their fine-grained approach leads to substantially higher energy savings than the previous coarse-grained approaches (primarily for memory-intensive workloads), they note that their proposed DVFS mechanism is also orthogonal to existing coarse-grained DVFS policies.

Schoonhoven et al. provide one of the first DVFS-capable auto-tuning tools for tuning GPU kernels~\cite{schoonhoven2022going}. Their approach extends the Kernel Tuner~\cite{kerneltuner} tool, which traditionally tunes for performance by testing different kernel parameters (thread block dimensions, loop unrolling, etc.). They show that although frequency is the dominant factor in deciding the energy usage, tuning for kernel parameters first and then tuning frequencies leads to different (Pareto-optimal) energy-efficient configurations than the other way around.

\paragraph{Inference + DVFS}
Nabavinejad et al. designed a runtime system that dynamically adjusts both the batch size and the frequency for DNN inference in (flexible) power-constrained systems~\cite{batchdvfs}. They find that the absolute range of power consumption the system can achieve through DVFS depends, through the utilization, on the batch size -- meaning that the lowest power consumption can only be achieved by lowering both. Their granularity for profiling the power and potentially switching the frequency is beyond 3 seconds.

Kakolyris et al. present a runtime system for LLM inference serving that applies DVFS to save energy for requests with service-level-objectives -- a.k.a. time budgets~\cite{10540202}. Like our work, they aim to reduce the energy consumption with a fixed performance loss guarantee (relaxed \textit{waste}). Using a pre-trained power and performance model, they apply the appropriate frequency at an iteration-level. They further highlight the challenge of applying DVFS in a highly variable inference environment where the requests rate, the batch size, and the input \& output size of each request can change continuously. DynamoLLM~\cite{10946802} pursues a similar strategy in a similar setting. Besides frequency, they also scale the number of instances and the tensor parallelism degree based on the current inference load. Frequencies are switched at a 5 second granularity.

\paragraph{Training + DVFS}
Perseus is a tool for reducing the energy waste when using pipeline parallelism in LLM training~\cite{perseus}. It saves energy by aggressively lowering the frequency (thereby gaining time) during idle stages (e.g., the pipeline bubble) or when one of the (other) pipelines encounters unusual delays (on the critical path) which increases the end-to-end latency. During training, the tool anticipates straggling pipelines based on IO and system throttling events and switches the other pipelines to a different (pass-granularity) frequency schedule. Their work is mostly orthogonal to ours, as we focus on (but are not limited to) a single pipeline setting with no idle stages.

Wang et al. present a fine-grained (operator-level) analysis of DVFS for LLM training using Huawei Ascend NPUs~\cite{10.1145/3669940.3707231} -- AI accelerators consisting of tensor cores. Through performance and power models, and using a genetic algorithm-based search, they are able to find clocks that reduce total energy consumption (of the compute units) by $13.44\%$. Through their use of NPUs, they are able to achieve $1$ millisecond switching latencies. Their experiments also show a higher switching latency can significantly decrease the potential energy savings. While their analysis is tailored to NPUs, our work focuses on GPUs, which exhibit substantially different DVFS mechanisms.


\section{Conclusion}
To accommodate AI compute growth, energy-efficiency optimizations must reduce energy consumption, without affecting performance, while being (to the user) low-effort, low-cost, and reliable. Through a case-study on training GPT-3, we show that DVFS can be applied such that it reduces \textit{compute waste}. Specifically, we argue that DVFS for LLMs needs a finer granularity and a more realistic optimization goal. For the former, we show that kernel-level DVFS could increase energy-efficiency, leading to a maximum of $15.64\%$ energy savings. 
For the latter, we show that optimizing for \textit{reducing compute waste} can determine effective clock configurations which reduce energy consumption without compromising performance -- demonstrating a pure power- and energy-efficiency improvement. 
We further show that these clock configurations perform similarly for different levels of data- and tensor parallelization -- with relative time and energy savings diverging by at most $2$ and $6$ percentage points, respectively. Thus, the savings will likely persist at scale.

Based on this empirical study, we foresee that a similar strategy could be applied to inference as well, though the search space would naturally grow. However, because the layer structure and their performance–energy characteristics remain consistent across inference runs, we expect the overall effort to remain comparable to prior pass- or iteration-level approaches. Furthermore, given that low DVFS switching latencies have already been demonstrated~\cite{predictdontreact}, we anticipate that finer-grained control -- such as kernel-level, or possibly even finer-level, DVFS -- will further increase the potential improvements.

Overall, we emphasize that even modest percentage-level energy savings can translate into substantial absolute reductions, given the massive scale of current and future LLM training and deployment infrastructures.



\bibliographystyle{ACM-Reference-Format}
\bibliography{ICPE_references}

\end{document}